\documentclass[10pt,letterpaper]{article}
\usepackage[top=0.85in,left=2.75in,footskip=0.75in]{geometry}

\usepackage{amsmath,amssymb}

\usepackage{changepage}

\usepackage[utf8x]{inputenc}

\usepackage{textcomp,marvosym}

\usepackage{cite}

\usepackage[hyphens]{url}
\usepackage{nameref,hyperref}

\usepackage[right]{lineno}

\usepackage{microtype}
\DisableLigatures[f]{encoding = *, family = * }

\usepackage[table]{xcolor}

\usepackage{array}

\newcolumntype{+}{!{\vrule width 2pt}}

\newlength\savedwidth

\newcommand\thickhline{\noalign{\global\savedwidth\arrayrulewidth\global\arrayrulewidth 2pt}%
\hline
\noalign{\global\arrayrulewidth\savedwidth}}

\raggedright
\usepackage[aboveskip=1pt,labelfont=bf,labelsep=period,justification=raggedright,singlelinecheck=off]{caption}

\makeatletter
\renewcommand{\@biblabel}[1]{\quad#1.}
\makeatother

\usepackage{lastpage,fancyhdr,graphicx}
\usepackage{epstopdf}
\fancyheadoffset[L]{2.25in}
\fancyfootoffset[L]{2.25in}
\usepackage{bm}

\usepackage{siunitx}
\usepackage{nicefrac}
\usepackage{color}

\usepackage{amsmath}
\renewcommand\d{\mathop{}\!\mathrm{d}}

\begin{document}
\vspace*{0.2in}
\begin{flushleft}
{\Large
\textbf\newline{{\it QFlow lite} dataset: A machine-learning approach to the charge states in quantum dot experiments} 
}
\newline
\\
Justyna P. Zwolak\textsuperscript{1,2,*}, 
Sandesh S. Kalantre\textsuperscript{1,2,3},
Xingyao Wu\textsuperscript{1,4},
Stephen Ragole\textsuperscript{1,4},
Jacob M. Taylor\textsuperscript{1,2,4}
\\
\bigskip

\textbf{1} Joint Center for Quantum Information and Computer Science, 
University of Maryland, College Park, MD, 20742, USA
\\
\textbf{2} National Institute of Standards and Technology, Gaithersburg, MD, 20899, USA
\\
\textbf{3} Department of Physics, Indian Institute of Technology - Bombay, Mumbai, 400076, India
\\
\textbf{4} Joint Quantum Institute, University of Maryland, College Park, MD, 20742, USA
\\
\bigskip

* jpzwolak@gmail.com
\end{flushleft}

\section*{Abstract}

Over the past decade, machine learning techniques have revolutionized how research and science are done, from designing new materials and predicting their properties to data mining and analysis to assisting drug discovery to advancing cybersecurity. Recently, we added to this list by showing how a machine learning algorithm (a so-called learner) combined with an optimization routine can assist experimental efforts in the realm of tuning semiconductor quantum dot (QD) devices. Among other applications, semiconductor quantum dots are a candidate system for building quantum computers. In order to employ QDs, one needs to tune the devices into a desirable configuration suitable for quantum computing. While current experiments adjust the control parameters heuristically, such an approach does not scale with the increasing size of the quantum dot arrays required for even near-term quantum computing demonstrations. Establishing a reliable protocol for tuning QD devices that does not rely on the gross-scale heuristics developed by experimentalists is thus of great importance. 

To implement the machine learning-based approach, we constructed a dataset of simulated QD device characteristics, such as the conductance and the charge sensor response versus the applied electrostatic gate voltages. The gate voltages are the experimental `knobs' for tuning the device into useful regimes. Here, we describe the methodology for generating the dataset, as well as its validation in training convolutional neural networks. From 200 training sets sampled randomly from the full dataset, we show that the learner's accuracy in recognizing the state of a device is $\approx$ 96.5~\% when using either current-based or charge-sensor-based training. The spread in accuracy over our 200 training sets is 0.5~\% and 1.8~\% for current- and charge-sensor-based data, respectively. In addition, we also introduce a tool that enables other researchers to use this approach for further research: {\it QFlow lite} -- a Python-based mini-software suite that uses the dataset to train neural networks to recognize the state of a device and differentiate between states in experimental data. This work gives the definitive reference for the new dataset that will help enable researchers to use it in their experiments or to develop new machine learning approaches and concepts.


\section*{Introduction}\label{sec:intro}
Quantum information---a field at the intersection of physics, mathematics, and computer science---seeks to harness the power of quantum systems for computing and communication technologies~\cite{Ladd10-QC,Nielsen11-QCI}. This field is currently undergoing rapid development as a series of candidate quantum computing platforms grow in scale and complexity~\cite{Gambetta17-QSC,Kielpinski02-LIC,Lekitsch17-MTQ,Zajac16-SGA,Mukhopadhyay18-2DD}. Essentially, coupled qubit systems---the quantum analog of classical bits---are moving from being a lab curiosity to fully integrated and usable computing devices. We are reaching the point where the uniquely quantum features of entanglement and superposition will enable a solution to problems that are thought intractable to even the most powerful present day supercomputers~\cite{Nielsen11-QCI}. 

This move, though, comes with a host of challenges that need to be overcome both for near term and long term computing goals. One of the greatest is the fabrication of stable, controllable, and scalable qubit arrays. There are a myriad of candidate systems to realize qubits---the fundamental building block of quantum computers---each having its own set of advantages and disadvantages to understand, improve, and control their operation~\cite{Li17-SQD,Karzig2017,Neill2017,Zajac16-SGA,Saffman2016,Sete2016,Blais2004,Brown2016,Bernien17}.

Electrons confined in semiconductor nanostructures, called quantum dots (QDs), are one such candidate. Quantum dots are defined by electrostatically confining electrons in nanowires or in two-dimensional electron gases (2DEGs) present at the interface of semiconductor heterostructures~\cite{Hanson07-SFQ}. In practice, for the linear dot arrays in a 2DEG, the electron density islands are generated in one-dimensional (1D) channels patterned in the 2DEG~\cite{Zajac16-SGA} thus allowing such channels to be modelled as effective nanowires, with electron transport restricted to 1D.

\begin{figure}[!t]
\includegraphics[width=1.0\linewidth]{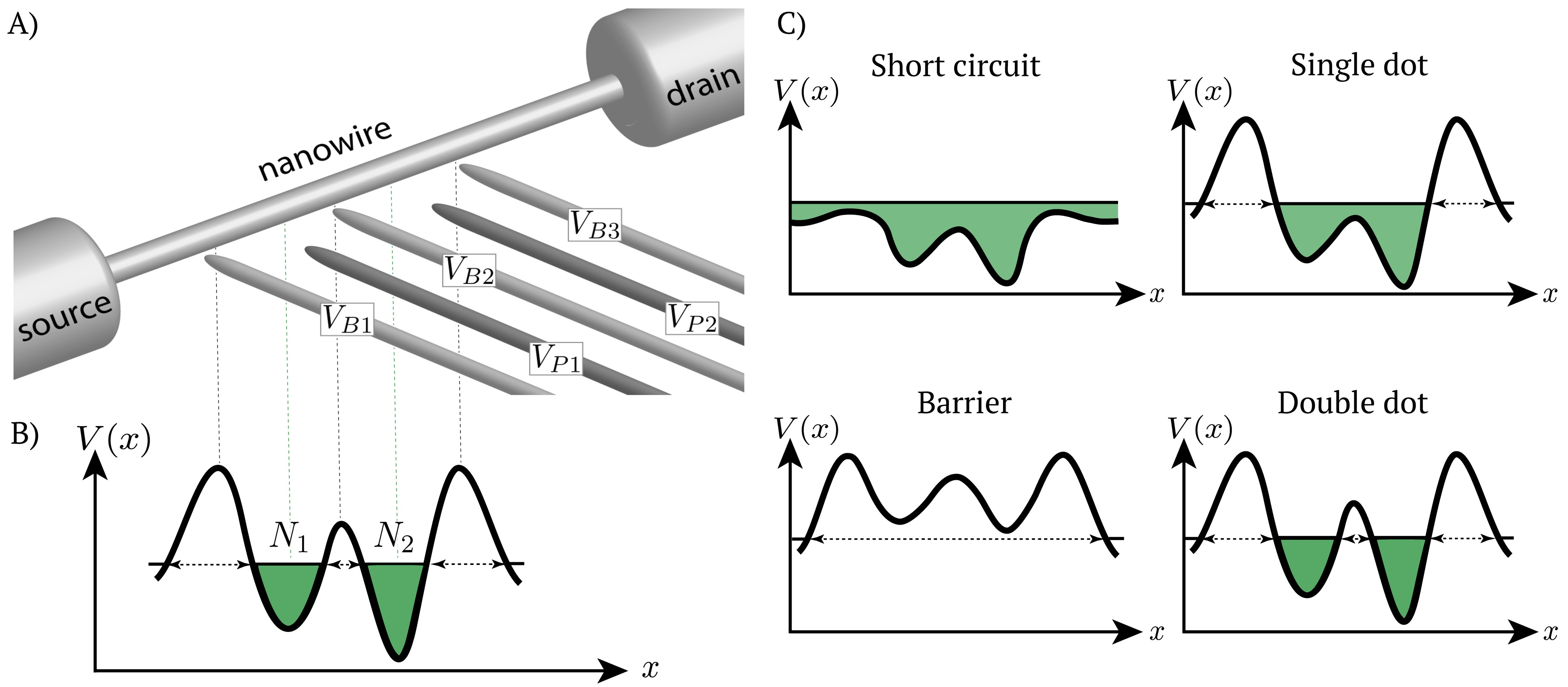}
\caption{{\bf Quantum dots from a nanowire.}
A) A generic model of a nanowire with 5 gates. The barrier gates, $V_{Bi}$ with $i=1,2,3$ (light gray), are set to a fixed voltage and are used to form islands by confining electron density to certain region. Voltage on the plunger gates, $V_{Pj}$ with $j=1,2$ (dark gray), is varied to allow for control of the current flow through the nanowire. B) Potential profile along a nanowire for a double dot system with $N_1$ and $N_2$ denoting the number of electrons on each dot. C)  Possible states in the 5-gate device. In the short circuit state the potential profile is below the Fermi level, leading to an unintended current flow. When the potential profile is above the Fermi level, the current flow is blocked (barrier state). By varying the voltage applied to plunger gates in the lower range while keeping the barriers above the Fermi level of the contacts, one can transition between one and two dots.}
\label{fig:DD-device}
\end{figure}

For semiconductor-based methods, the realization of good qubit performance is achieved via electrostatic confinement, band-gap engineering, and dynamically adjusted voltages on nearby electrical gates. Fig~\ref{fig:DD-device} shows a generic setup of quantum dots in a nanowire. The voltages applied to barrier gates ($V_{Bi}$, $i=1,2,3$) and plunger gates ($V_{Pj}$, $j=1,2$) define the potential landscape in which the quantum dots are formed. In particular, barriers define the dot position by locally depleting carriers beneath them, thereby separating the electron density into disjoint regions, while the plungers shift the chemical potential in the dots relative to the chemical potentials of the contacts. In other words, the choice of these voltages determines the number of dots, their position, and their coupling, as well as the number of electrons present in each dot. 

The translation of requisite dot locations and charges to gate voltages presents a difficult classical control problem. Currently, gate voltages are set heuristically in order to reach a stable few electron configuration in experiment. However, given the recent progress in the physical construction of larger arrays of quantum dots~\cite{Zajac16-SGA,Mukhopadhyay18-2DD}, it is imperative to have a reliable and scalable method to find a stable, desirable electron configuration in order to initialize a quantum dot array. The ultimate goal is to have a fully automated method to find the voltages that yield the right number of dots, electrons, and couplings. There have been initial attempts to use computer-supported, algorithmic experimental device control, though these approaches typically still require human input and device-dependent parameters~\cite{Baart2016,Botzem2018,vanDiepen18-ATC}.

In recent years, machine learning (ML) algorithms, and specifically convolutional neural networks (CNNs), have emerged as a ``go to'' technique for automated classification, giving reliable output when trained on a representative, comprehensive dataset~\cite{Krizhevsky12-CCN}. A natural next step towards a fully automated tuning of experiments is to combine the computer-supported device control with machine learning techniques to replace the heuristics developed by experimentalists~\cite{Wiel02-DQD}. A new ``auto-tuning'' paradigm, where these two approaches are put together, has recently been realized in the context of single and double dot devices~\cite{Kalantre18-MTQ}. In their work, Kalantre {\it et al.} proposed a closed-loop system of experimental control using a large, labeled simulated dataset, CNN-based learner, and numerical optimization techniques to automatically tune the quantum dot device.

As with all ML techniques, the quality of the training data is essential for successful implementation of the learner. In scenarios like quantum dot tuning, the availability of a physical model that can qualitatively mimic experimental output is key to providing large, reliable datasets for training. In our case, we developed such a dataset of simulated current, number of charges on each dot, sensor response, and state identifiers, as a function of the gate voltages. This dataset was employed to train a neural network and design an auto-tuning protocol that was further evaluated on simulated and experimental data~\cite{Kalantre18-MTQ}. Moreover, due to the paramount importance of good quality training data, we released our dataset for future machine learning studies via the Midas system at National Institute of Standards and Technology (NIST) and at data.gov~\cite{qf-data}.

Here, we describe the generation and validation of the dataset employed in Ref.~\cite{Kalantre18-MTQ}, along with a bird's-eye view of auto-tuning of quantum dot devices. While the physics of the model we use is presented in detail in Appendix A in Ref.~\cite{Kalantre18-MTQ}, we include here a brief overview for completeness and to set the foundations for the charge sensor model. We also introduce {\it QFlow lite} -- a Python-based mini-software suite that uses the dataset to train neural networks to recognize the state of a device and differentiate between states in experimental data. This work thus gives the definitive reference for the dataset (which can be found at Ref.~\cite{qf-data}) to enable researchers to use it in their experiments, as well as in the development of new ML approaches and concepts.

\section*{Materials and methods}
To mimic the transport characteristics of an experimental device, we developed a model for electron transport in gate-defined quantum dots (again, see Appendix A in Ref.~\cite{Kalantre18-MTQ} for additional information). Here we extend it to include also the charge sensor data generation. Note that this model is at best a qualitatively correct approach; the general interacting fermionic problem requires a quantum computer to fully solve, though we expect that the quasi-1D nature of these devices enables (a computationally very expensive) classical simulation to do somewhat better than our inexpensive, modified Thomas-Fermi solver. 

A generic gate-defined double dot device is presented in Fig~\ref{fig:DD-device}A. The barrier gate electrodes, labeled in Fig~\ref{fig:DD-device}A as $V_{Bi}$ ($i=1,2,3$), are used to confine electron density to certain regions, forming islands of electrons. The plunger gates, labeled $V_{Pj}$ ($j=1,2$), shift the chemical potential in a given dot relative to the chemical potentials of the contacts (reservoirs of electrons, labeled in Fig~\ref{fig:DD-device}A as the source and drain) that are assumed to be kept at a fixed chemical potential. In our simulations, the electron islands are generated in a 1D channel designed in a 2DEG.

\begin{figure}[!t]
\includegraphics[width=0.9\linewidth]{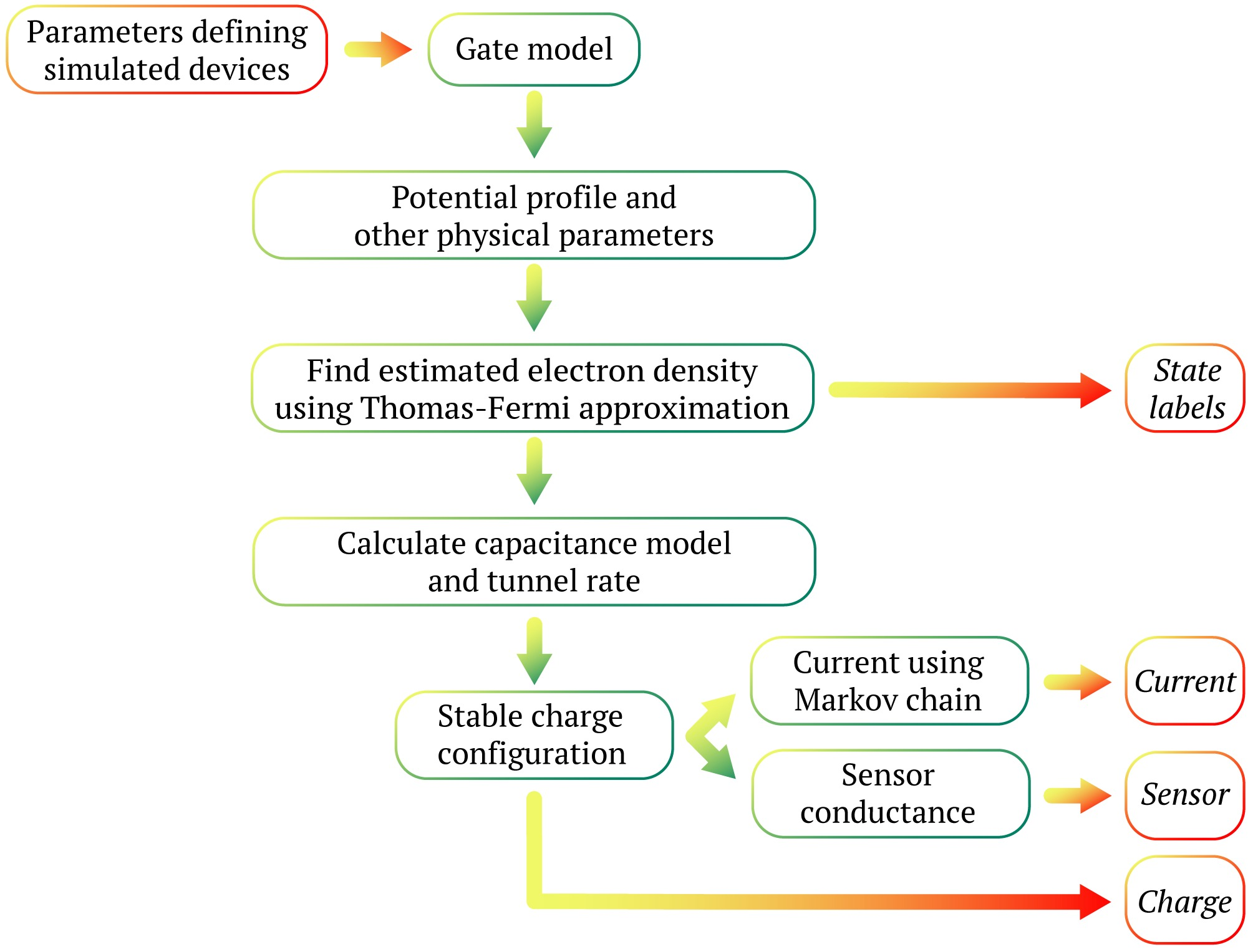}
\caption{{\bf Data simulation flow chart.} The simulation for a device begins by setting gate voltages and calculating the potential profile $V(x)$. This potential profile, along with other physical parameters, is used to calculate the electron-density self-consistently along the 1D channel. The state of the channel, e.g., the number of dots is known at this stage. The electron density is used to construct a capacitance model. The model predicts the stable charge states on the dots and the current through the device. The sensor conductance is calculated using the charge states. The final output for a single set of gate voltages consists of the device state (state labels), current, charges, and sensor conductance(s). The simulation is repeated for every point in the space of plunger gate voltages for a single device and then across an ensemble of device geometries.}
\label{fig:sim-flow}
\end{figure}

Applying voltages to the gates defines a potential profile along the 1D channel, $V(x)$. Depending on the relation between the chemical potential $\mu$ and the electrostatic potential $V(x)$, alternating regions of electron density islands and barriers are formed (see Fig~\ref{fig:DD-device}B), i.e., the state of the device is established. For the 5-gate device the possible states are a barrier or quantum point contact (QPC), single dot (SD), double dot (DD), and short circuit (SC) (see Fig~\ref{fig:DD-device}C). To go from the potential profile to experimentally measurable characteristics of the device, we use a physical model described in the following section. The flow of the data simulation is visualized in Fig~\ref{fig:sim-flow}. 

\subsection*{Physical model of the device}\label{sec:physical-model}
In our model, we use the theory of many-electron systems developed by Thomas and Fermi~\cite{Lundqvist83}, treating the electronic density $n(x)$ as the main variable. While such an approach lacks the quantitative details of a full simulation, it reproduces the minimum qualitative features of the system sufficient for machine learning, as we will show. 

\subsubsection*{Electron transport}
As mentioned earlier, in the transport simulations, we assume that the externally created 1D potential $V(x)$ is given. In particular, the barrier voltages are set to $V_{Bi} = -200\,$\si{\milli \volt}, for $i=1,2,3$. For plungers, voltages vary between $0\,$\si{\milli \volt} and $400\,$\si{\milli \volt}. The ends of the device are connected to reservoirs of electrons which are assumed to be kept at a fixed chemical potential, $\mu_1, \mu_2 \approx 100\,$\si{\milli \electronvolt}, with a very small bias of $100\,$\si{\micro \electronvolt} present across the leads.

The electron density, $n(x)$, is found by iteratively solving the self-consistent equation:
\begin{equation}\label{eqn:elec-density}
    n = \int_{\epsilon_0}^{\infty} \frac{g(\epsilon)}{1 + e^{\beta(\epsilon - E_F)}} \d\epsilon, 
\end{equation}
with the starting solution $n(x) = 0$. Here, $\beta$ is the inverse temperature, $\epsilon_0$ is the conduction band minimum, and $g(\epsilon)= g_0 = \frac{m*}{\pi \hbar^2}$ is the density of states in the conduction band set to a constant $g_0$ for a 2DEG. The conduction band minimum itself is modified because of the presence of the electron density, hence the need for self-consistency. The modified band minimum is calculated as
\begin{equation}\label{eqn:mod-band-min}
    \epsilon_0'(x) = \epsilon_0 - e V(x) + \int K(x,x') n(x') \d x',
\end{equation}
where $\epsilon_0'(x)$ is the new spatially varying band minimum, $V(x)$ is the externally applied potential, and $K(x,x') = \frac{K_0}{\sqrt{(x-x')^2 + \sigma^2}}$ gives the Coulomb potential between points $x$ and $x'$. $K_0$ sets the energy scale of the interaction and $\int K(x,x') n(x') \d x'$ gives the effective Coulomb potential created as a result of the electron density $n(x)$. A softening parameter $\sigma$ has been added to the denominator to help model the effective 1D interaction for a higher dimensional gas of electrons as would be present in the device and to prevent a numerical singularity at $x=x'$. The modified band minimum is in turn used to find the $n(x)$, and the process is repeated until the density $n(x)$ converges. The strength of the Coulomb interaction is being increased in a linear fashion to the required level for a fixed initial number of iterations to avoid pathologies associated with numerical convergence in the self-consistent calculation.

\subsubsection*{Capacitance model and tunnel rates}
Once the electron density is determined, we cast the system of islands into a capacitance model~\cite{Wiel02-DQD}. In such a model, each island of electron density is coupled to the depletion gates and other islands via capacitors. An integer number of electrons are assumed to be present on each island. Let $\bm{Q}$ be the vector containing the integer charges on the islands and $\bm{Z}$ be the induced charge from the depletion gates. The energy of such a capacitance model is given as:
 \begin{align}\label{eqn:cap-model}
 E &= (\bm{Q}-\bm{Z})^{T}\left(\frac{1}{2 \mathbb{C}}\right)^{-1}(\bm{Q}-\bm{Z}) \\
   &= \sum_{i,j} E_{i,j} (\bm{Q}-\bm{Z})_i (\bm{Q}-\bm{Z})_j,
 \end{align}
where $\mathbb{C}$ is the capacitance matrix and ${E_{i,j}} = \left(\frac{1}{2\mathcal{C}}\right)^{-1}_{i,j}$.

The objective behind constructing a capacitance model is to find the stable configuration of charges $\bm{Q}$ that leads to the minimum energy $E$. The Thomas-Fermi electron density $n(x)$ gives an estimate of the inverse capacitance elements as
\begin{align}
E_{i,j} = \frac{c_k \delta_{i,j} \int_{i}n(x)^2 \d x + \int_{i}\int_{j}K(x,x') n(x)n(x') \d x}{\left( \int_{i}n(x) \d x\right)\left( \int_{j}n(x) \d x\right) },
\end{align}
where $ \delta_{i,j}$ is the Kronecker delta function and $c_k$ is the coefficient that sets the scale for the kinetic energy of the Fermi sea on each island. The integration subscript $\int_i$ denotes that the integration is to be performed only over the $i^{\text{th}}$ island. The denominator normalizes to the total number of electrons on each island.

The islands of electron density are separated by potential barriers. Quantum mechanically, it is possible for electrons to tunnel into adjacent islands or contacts. We calculate a tunneling rate for a single electron from the islands to one another or to the contacts under the WKB approximation. Tunnel rates are then used to set a Markov chain based on charge states to calculate the current. In this approach, we calculate the probability of the single electrons to tunnel to adjacent islands. The stable configuration of the Markov chain amongst the charge states allows us to calculate a current through the device (see Ref.~\cite{Kalantre18-MTQ}). 

\subsubsection*{Charge Sensor Model}
In our simulations, in addition to finding the current, we also calculate the response of a sensor to the charge on the islands (see Fig~\ref{fig:sim-flow}). A charge sensor is a device capable of sensing the electrostatic environment around it~\cite{Johnson05-PHD}. It can be constructed in two principal ways: using a quantum point contact (QPC) or using a single dot (SD). In both cases, the current through the sensor is sensitive to the change in potential at the sensor's position. A high sensitivity is achieved by tuning it to the rising flank of a conductance step (in the case of a quantum point contact) or a Coulomb blockade peak (in the case of a quantum dot). Generically, we model the output of a charge sensor with a simplified model. The sensor conductance, $g_s$, is given as the linear combination of the voltages produced at the sensor location by each of the quantum dots:
\begin{equation}\label{eqn:sensor-conductance}
    g_s \propto \sum_i \frac{Q_i}{r_i},
\end{equation}
where $Q_i$ and $r_i$ are the charge on and the distance of the sensor from the $i^{th}$ quantum dot, respectively. The position of the quantum dot is determined from the peak positions of the electron density islands produced from the hypothetical Thomas-Fermi model. In our dataset for the 5-gate device, we store the response produced by two charge sensors positioned in a 1D channel $50\,$\si{\nano \meter} away from the channel where quantum dots are created.

\subsubsection*{Gate Model of the Simulated Devices}
The gates in the simulated QD devices are modelled as cylindrical conductors. Each gate is defined by four parameters: $x_0$ (the gate position), $r_0$ (the radius of the gate), $h$ (the height from the 2DEG at which gates are fixed), and $V_0$ (the height of the potential profile at the gate position) that are randomly sampled from a Gaussian distribution with the mean values listed in Table~\ref{tab:5-gates} and standard deviation set to $0.05$ times the mean value. The total width of the device is set to $120\,$\si{\nano \meter}.  

\begin{table}[t]
\centering
\caption{
{\bf Mean values for the parameters defining simulated 5-gates devices: height of the potential profile ($V_0$) at the gate position ($x_0$), the height at which barriers were fixed ($h$), and the radius of the barrier gates ($r_0$). The device size along $x$-axis is $120\,$\si{\nano \meter}, with the center positioned at $x_0=0\,$\si{\nano \meter}.}}
\begin{tabular}{|l+l|l|l|l|}
\hline
   {\bf Gate} & {\bf$V_0$ (\si{\milli \volt})} & $x_0$ (\si{\nano \meter}) & $h$ (\si{\nano \meter}) & $r_0$ (\si{\nano \meter})  \\ \thickhline
    $B1$ & -200  & $-40$ & 50  & 5\\  \hline
    $P1$ & (0,400)  & $-20$ & 50 & 5 \\  \hline
    $B2$ & -200 & 0 & 50 & 5\\  \hline
    $P2$ & (0,400) & 20 & 50 & 5\\  \hline
    $B3$ & -200 & 40 & 50 & 5\\  \hline
\end{tabular}
\label{tab:5-gates}
\end{table}

The potential profile for each device is given as:
\begin{equation}
V(x) = \frac{V_0}{\log\nicefrac{h}{r_0}} \log{\frac{\sqrt{(x-x_0)^2 + h^2}}{r_0}} \exp{\left(-\frac{|x-x_0|}{\sigma_{sc}}\right)},
\end{equation}
where $V_0$ sets the height of the potential profile at $x = x_0$ and $h$ controls the width of the profile~\cite{Griffiths99}. The term $e^{-\frac{|x-x_0|}{\sigma_{sc}}}$, with $\sigma_{sc} = 20\,$\si{\nano \meter} equal to the separation between adjacent gates, has been added to account for the screening due to the electron density present in the semiconductor. 

Additionally, each device is characterized by a set of standard physical parameters chosen to assure that the quantum dots would be in the few electron (0 to 10 electrons) regime:
\begin{itemize}
\item $K_0 = 10\,$\si{\milli \electronvolt} -- sets the strength of the Coulomb interaction, 
\item $\sigma = 2\,$\si{\nano \meter} -- prevents blowup at $x=0$ in the interaction,
\item  $g_0 = 0.5\,$\si{\electronvolt}$^{-1}\,$\si{\nano \meter}$^{-1}$ -- sets the scale for the density of states, 
\item $c_k = 1\,$\si{\milli \electronvolt \nano \meter} -- gives the kinetic term for the 2DEG. 
\end{itemize}

The current, charge, and charge sensor response are calculated as a function of  $(V_{P1},V_{P2})$. The resultant dataset allows to train the neural network for a large range of potential experimental devices, creating a tuning procedure that is robust against varying conditions and imperfections.

\subsection*{Overview of the {\it QFlow lite} dataset}\label{sec:data-structure}
As described in previous section, we calculate the electron density in a reference semiconductor system comprising of a quasi-1D nanowire with a series of depletion gates  using a modified Thomas-Fermi approximation. Those gates voltages determine the number of dots and the charges on each of the dots, as well as the conductance through the channel. Using the sampling procedure for physical parameters and gate configurations described in the previous section, we generated an ensemble of 1001 different realizations of the same type of device. For each device realization, we stored data as a $100\times100$ map from plunger gate voltages to the specific output domains. In particular, we stored the current through the device at small bias and the charge sensor readout (the relevant, experimentally measurable quantities), as well as the number of charges on each island and the state labels (not directly measurable but useful for machine learning). The machine learning training and evaluation set is then generated by taking random sub-images of fixed size from each map.

The output from simulations, together with the plunger voltage ranges and all physical parameters defining a given device, are stored as NumPy data files, as this file format allows for storing structured data while keeping the size of a file relatively small (the total size of a compressed zip data file with the raw data is 913.1~MB)~\cite{scipy,numpy}. Moreover, NumPy files enable faster reading compared to plain text or CSV files, which is particularly important for high performance scientific computing and data analysis~\cite{numpy-performance}. The Jupyter notebook in the S1 Supporting information provides a Python code to convert the NumPy files to csv format. 

\begin{figure}
\includegraphics[width=0.9\linewidth]{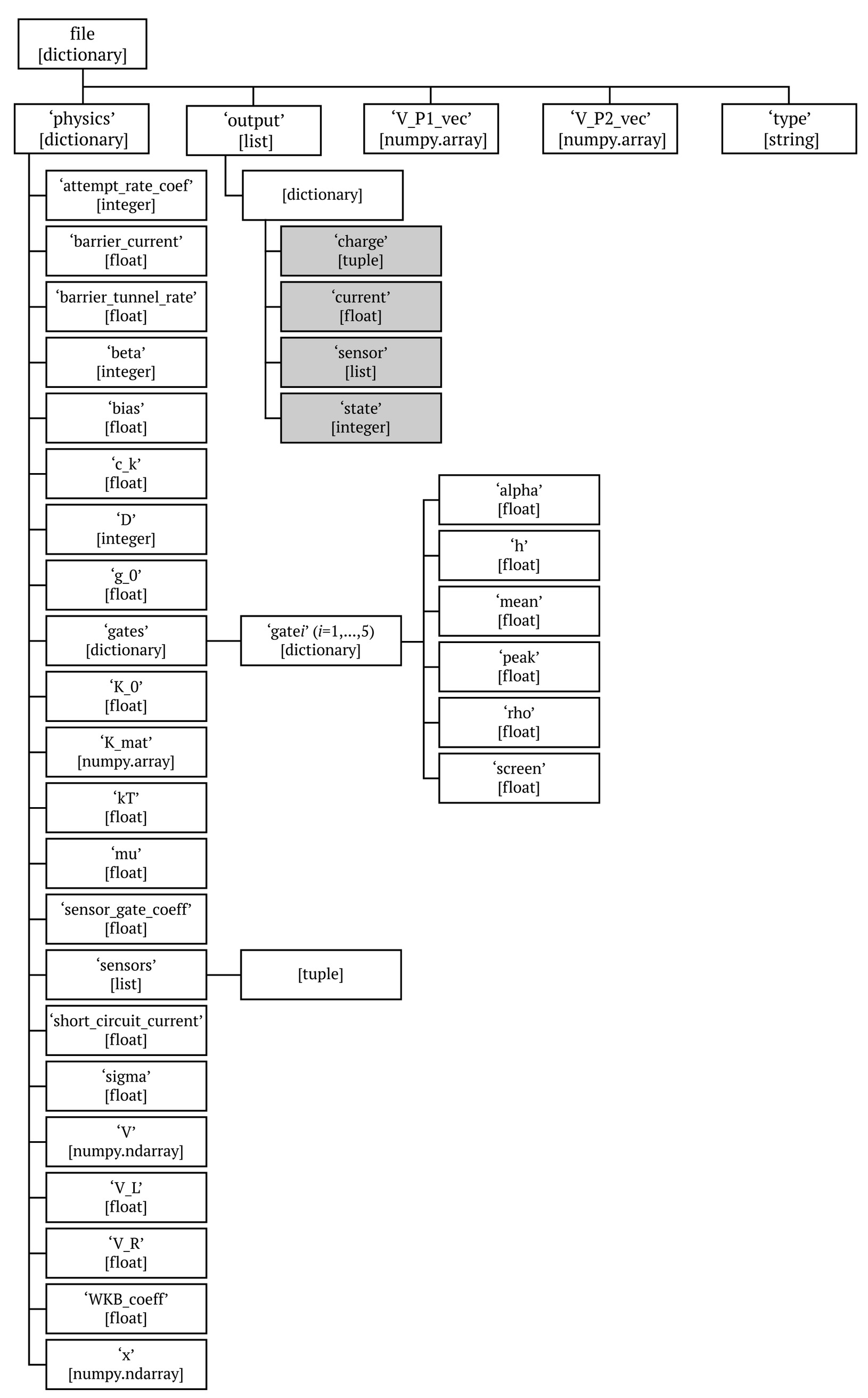}
\caption{{\bf Data structure.}
The generic data structure tree for the data files. The data type is given in square brackets. The simulation output is highlighted in gray. See Table~\ref{tab:output} and Table~\ref{tab:physics} for a reference list.
}
\label{fig:tree-diagram}
\end{figure}

Each file contains information about a single simulated device. While the units are not included as part of the dataset, they are provided here for completeness. The generated devices are stored as dictionaries with the following five elements (keys) in them (the type of each element in the dictionary is given in the brackets):
\begin{itemize}
\item[] {\bf `type'}: information about what data is in the file [string],
\item[] {\bf `V\_P1\_vec'}: voltage range for the first plunger ($0$ to $0.4\,$\si{\volt}) [numpy.array], 
\item[] {\bf `V\_P2\_vec'}: voltage range for the second plunger ($0$ to $0.4\,$\si{\volt}) [numpy.array], 
\item[] {\bf `output'}: the simulated data [list];
\item[] {\bf `physics'}: physical parameters of the device [dictionary]; 
\end{itemize}
For the structure of full data files, see Fig~\ref{fig:tree-diagram}. For a detailed reference list for the simulated data (i.e., `output'), refer to Table~\ref{tab:output}, and for a reference list for the physical parameters of the device (i.e., `physics'), see Table~\ref{tab:physics}.

\begin{table}[t]
\centering
\caption{
{\bf The single device simulation output (stored as `output') is a list of 10\,000 dictionaries, holding the simulated data for each point in the plunger voltage space (defined by vectors `V\_P1\_vec' and `V\_P2\_vec'). The `output' dictionaries include four variables, as defined in the table.}}
\begin{tabular}{|l+p{0.72\linewidth}|l|}
\hline
{\bf Key} & {\bf Description} & {\bf Type}  \\ \thickhline
`charge' & an information about the number of charges on each dot (with a default value 0 for short circuit and a barrier) & tuple \\ \hline
`current' & a current through the device at infinitesimal bias & float\\ \hline
`sensor' & an output of the charge sensors, evaluated as the electrostatic potential at the sensor locations & list \\ \hline
`state' & a numeric label determining the state of the device, distinguishing between a single dot (1), a double dot (2), a short circuit (-1), and a barrier (0) & integer \\ \hline
\end{tabular}
\label{tab:output}
\end{table}

\begin{table}[!ht]
\begin{adjustwidth}{-2.25in}{0in} 
\centering
\caption{
{\bf The physical parameters of the devices, stored as a dictionary `physics'. Fixed values are given explicitly. Varied parameters, given in angle brackets, were randomly sampled from a Gaussian distribution with the given mean value $\mu$ and standard deviation set to $0.05\,|\mu|$ (unless stated otherwise).}}
\begin{tabular}{|p{0.16\linewidth}+p{0.57\linewidth}|p{0.18\linewidth}|}
\hline
{\bf Key} & {\bf Description} & {\bf Value}  \\ \thickhline
attempt\_rate\_coef & controls the strength of the attempt rate factor & $1$ \\ \hline
barrier\_current & a scale for the current set to the device when in barrier mode & $1$ arb. unit \\ \hline
barrier\_tunnel\_rate & a tunnel rate set when the device is in barrier mode while calculating the tunnel probability & $10.0$ \\  \hline
beta & effective temperature used for self-consistent calculation of the electron density $n(x)$ & $1000\,(\si{\electronvolt})^{-1}$ \\ \hline
bias & difference in the chemical potential between the source and drain & $100\,\si{\micro \electronvolt}$ \\ \hline
c\_k & kinetic term for the 2DEG & $\langle1\,\si{\milli \electronvolt \nano \meter}\rangle$ \\  \hline
D & dimension of the problem to be used in the electron density integral, (only when polylogarithm function is used to calculate the electron density, for a 2DEG a direct analytic integral of the Fermi function is used) & 2 \\  \hline
g\_0 & coefficient of the density of states & $\langle 1.0\,(\si{\electronvolt \nano\meter})^{-1}\rangle$ \\ \hline
gates & the dictionary of parameters defining each of the five gates: & \\  
 &  {\it alpha}: lever arm (same for all gates) & $\langle 1.0 \rangle$\\
 &  {\it h}: distance of the gate from the electron density (same for all gates) & $\langle 50.0\,\si{\nano \meter} \rangle$\\
 &  {\it mean}: position of the gate along linear array & \\
  &   - for gate 1 & $\langle -40\,\si{\nano \meter} \rangle$ \\
  &   - for gate 2 & $\langle -20\,\si{\nano \meter} \rangle$ \\
  &   - for gate 3 & $\langle  0\,\si{\nano \meter} \rangle$ \\
  &   - for gate 4 & $\langle 20\,\si{\nano \meter} \rangle$ \\
  &   - for gate 5 & $\langle 40\,\si{\nano \meter} \rangle$ \\
 &  {\it peak}: potential at the location of the electrons&  \\
 &   - for gates 1, 3, and 5 & $\langle 200\,\si{\milli \volt}\rangle$ \\
 &   - for gates 2 and 4 & $\langle -400\,\si{\milli \volt}\rangle$ \\
 &  {\it rho}: radius of the cylindrical gate (same for all gates) & $\langle 5.0\,\si{\nano \meter} \rangle$\\ 
 &  {\it screen}: the screening length (same for all gates) & $\langle 20.0\,\si{\nano \meter}\rangle$ \\ \hline
K\_0 & the strength of the Coulomb interaction & $\langle10\,\si{\milli \electronvolt}\rangle$ \\ \hline
K\_mat & the Coulomb interaction matrix & K\_mat(x,K\_0,sigma)\\  \hline 
kT & temperature of the system used in the transport calculations & $50\,$\si{\micro \electronvolt} \\ \hline
mu & Fermi level (assumed to be equal for both leads) & $0.1\si{\electronvolt}\,$ \\ \hline
sensor\_gate\_coeff & weight applied while including the potential of the gate in calculating the sensor output & $0.1$ \\ \hline
sensors & the position of the two charge sensors in the 2DEG plane, stored as (horizontal position with respect to the center of the device, vertical position with respect to the dots which are assumed to be located on the $x$-axis) & $[(-20, 50), (20, 50)]\,\si{\nano \meter}$ \\  \hline
short\_circuit\_current & an arbitrary high current value given to the device when in short circuit mode & $100$ arb. unit \\  \hline
sigma & softening parameter & $3.0\,\si{\nano \meter}$ \\  \hline
V  & potential profile & V(x) \\ \hline
V\_L & voltage applied to left lead & $50\,\si{\micro \volt}$ \\  \hline
V\_R & voltage applied to right lead & $-50\,\si{\micro \volt}$ \\  \hline
WKB\_coeff & the strength of WKB tunneling & $0.5$ \\ \hline
x & linear array spanning the size of the device & $(-60,60)$\,\si{\nano \meter} \\ \hline 
\end{tabular}
\label{tab:physics}
\end{adjustwidth}
\end{table}

\subsection*{Data visualization}\label{sec:data-visual}
\begin{figure}[!t]
\includegraphics[width=1.0\linewidth]{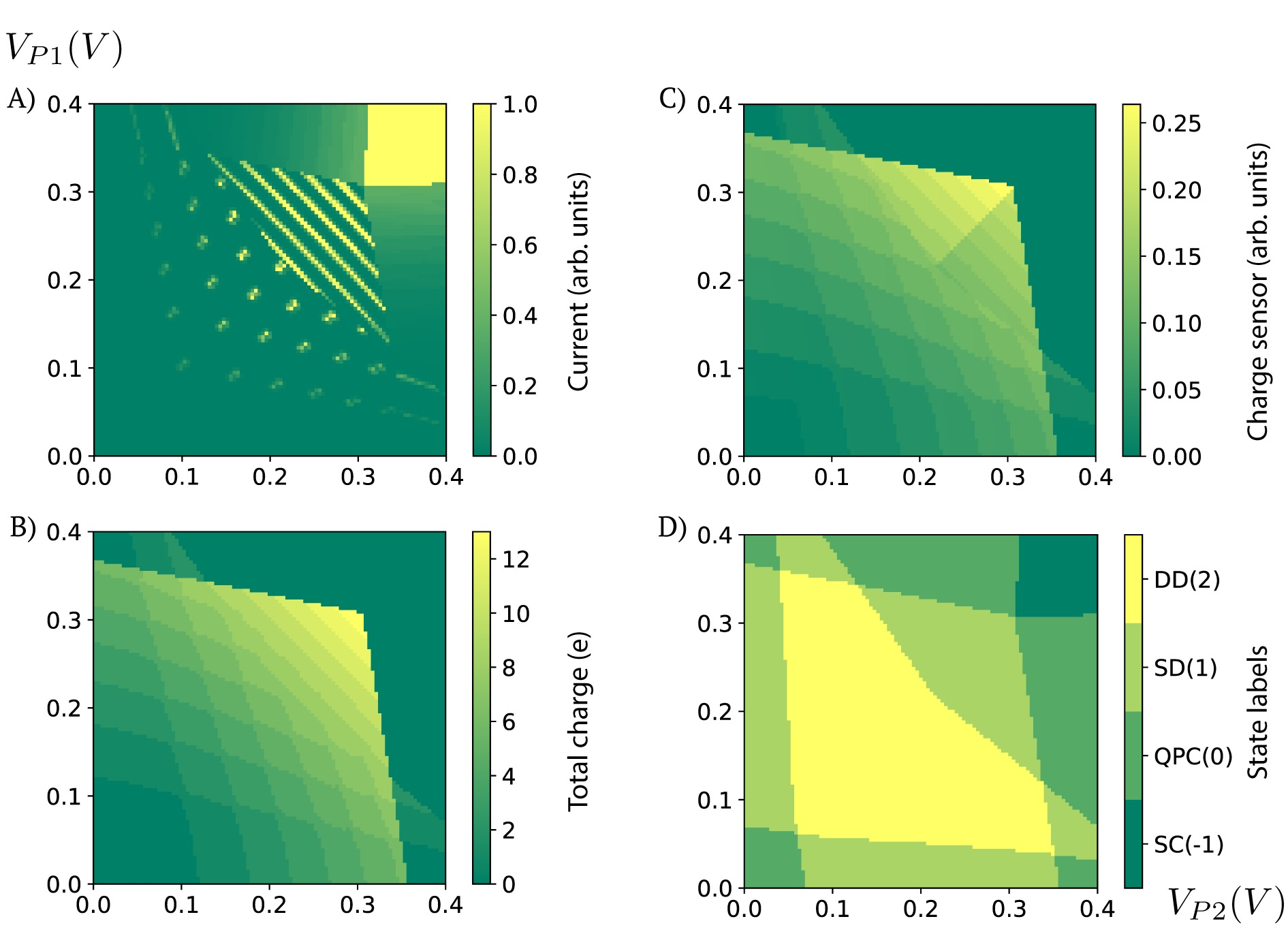}
\caption{{\bf Data visualization.}
A) Current, B) charge, C) charge sensor, and D) state data as a function of plunger gate voltages (2D map 100 $\times$ 100 pixels). Note that the data is unitless. For current, the data in the figure is re-scaled by a factor of $10^4$. The code used to generate these plots can be found in S1~Supporting information.} 
\label{fig:data-visual}
\end{figure}

The data stored in the `output' correspond to a 2D map ($100\times100$ pixel) from the space of plunger gate voltages ($V_{P1}$, $V_{P2}$) to current, charge, state, and sensor maps (each stored separately). These maps can be used as observables (current, charge, sensor) and labels (state) for machine learning training and testing. Figure~\ref{fig:data-visual} shows a preview of a sample of each of the generated data types. The Python code that allows for visualizing the `output' data is included in the Jupyter notebook (S1 Supporting information).

\section*{Results and discussion}
\subsection*{Dataset validation}
To ensure that the dataset can be used by independent researchers to generate their own training sets, we performed a validation test. Specifically, to test the reliability of the dataset, we randomly generated 200 full training sets of $10,010$ subregions ($30\times30$ pixels) that were then used for training and evaluation. These subregions were used as the input to the convolutional neural network. To mimic the experimental setting, for charge sensor data we considered the gradient of the simulated data (i.e., the differential conductance) rather than the raw data. We then run five iterations of network training per full set, each time dividing it into two disjoint parts with 90 \% of images used for training and 10 \% of images used for  evaluation. The network used for training is comprised of one convolutional layer with 16 filters of the size $5\times5$, followed by one pooling layer and three dense layers, with 1024, 512, and 128 units, respectively. By convolving the input data (i.e., ``sliding'' the filters over the image and computing the dot product of the filter with the input data), the convolutional layer extracts high-level features from the input image while preserving the spatial relationship between pixels. The pooling layer reduces the size of each feature map while retaining the most important information. Finally, the dense layers allow for learning non-linear combinations of these features and to classify the data. We used the Adam optimizer~\cite{Diederik14} with a learning rate $\eta=0.001$, 5000 steps per training, and a batch size of 50. 

\begin{figure}[!t]
\includegraphics[width=\linewidth]{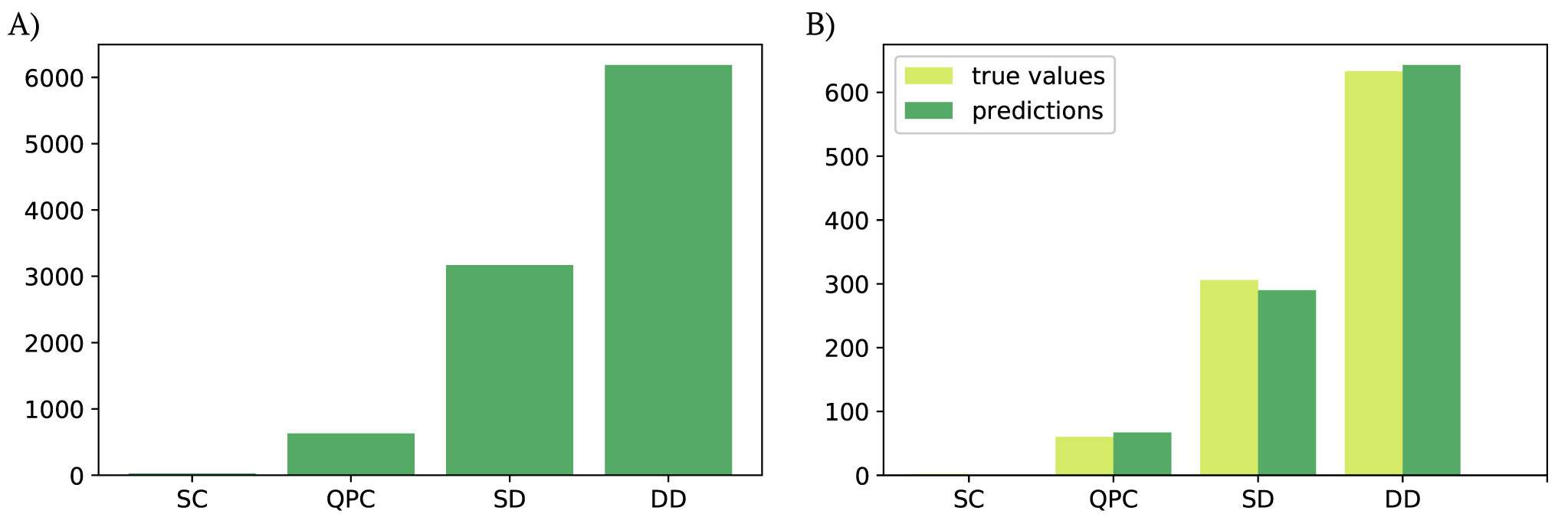}
\caption{{\bf State distribution.}
A) A typical distribution of states in a sample training set ($N=9009$). B) A visualization of the performance of the ML algorithm on sample simulated data ($N=1001$).} 
\label{fig:ml-dist}
\end{figure}

The accuracy of the learner is defined as a percentage of correctly identified images (supervised learning). To account for the uneven distribution of representatives between classes in the training sets (see Fig~\ref{fig:ml-dist}A for a typical distribution of the states in the training set), we used a weighted average of F1 scores of each class to find accuracy for each iteration~\cite{Chinchor92-F1}. Note that in our tests, we did not focus on achieving a high recognition accuracy but rather on verifying the consistency of the accuracy rate between iterations. The overall average accuracy of the learner for the current data is 96.3 \%, with a standard deviation (std dev) of 0.5 \% and for the charge sensor data the average accuracy is 95.9 \% (std dev = 0.6 \%). The training takes about 4.5 min (std dev = 1.1 min) for the current data and about 4.6 min (std dev = 0.2 min) for the charge sensor data on a 2014 MacBook Pro.

The average accuracy per state type, i.e., the percentage of correctly identified SD, DD, etc., is in both cases fairly comparable for all states except for SC. For the current data, the average accuracy is 63.5 \% (std dev = 29.2 \%) for SC, 92.8 \% (std dev = 1.9 \%) for QPC, 94.5 \% (std dev = 0.8 \%) for SD, and 97.7 \% (std dev = 0.4 \%) for DD. For the charge sensor data, the average accuracy is 22.4 \% (std dev = 26.1 \%) for SC, 92.8 \% (std dev = 2.1 \%) for QPC, 94.0 \% (std dev = 0.8 \%) for SD, and 97.4 \% (std dev = 0.4 \%) for DD. Figure~\ref{fig:ml-dist}B shows visualization of the performance of a ML algorithm described earlier, per state, on a sample simulated current-based dataset. The lower accuracy for SC is caused by the significantly lower number of SC images in the training and evaluation sets. The parameters of the ML algorithm, such as the number of convolutional and pooling layers, the number and size of dense layers, the optimizer and learning rate, batch size and number of steps can be modified to further improve the performance of the trained network.

The goal of the project presented in Ref.~\cite{Kalantre18-MTQ} was to auto-tune the device to a DD configuration. Thus, the main task of the ML algorithm was to distinguish between SD and DD states. The low number of QPC and SC states does not affect performance of the trained network. In order to obtain a training set that includes more comparable numbers of all types of states one can, e.g., apply affine transformations on the underrepresented type to reach the desired states distribution. As has been shown in Ref.~\cite{Kalantre18-MTQ}, however, the network trained on the {\it QFlow lite} dataset can be employed to auto-tune simulated and, in principle, experimental devices.

\subsection*{{\it QFlow lite} suite: working with the data}\label{sec:qflow}
To ease working with the dataset, we developed a software suite---{\it QFlow lite}---that allows users to generate current-based training sets for machine learning based on the {\it QFlow lite} dataset, to modify the network's architecture, to train the network, and to test it on their own experimental data. {\it QFlow lite} is written in Python and operated through a Jupyter notebook-based interface~\cite{jupyter}. It can be accessed on GitHub~\cite{qf-lite}. The software comes with a ``User Guide'' that includes the software dependencies required for successful execution of the code, a manual explaining the workflow of {\it QFlow lite}, and guidance on how to convert the experimental data to a format compatible with the trained network. 

\begin{figure}[!t]
\includegraphics[width=\linewidth]{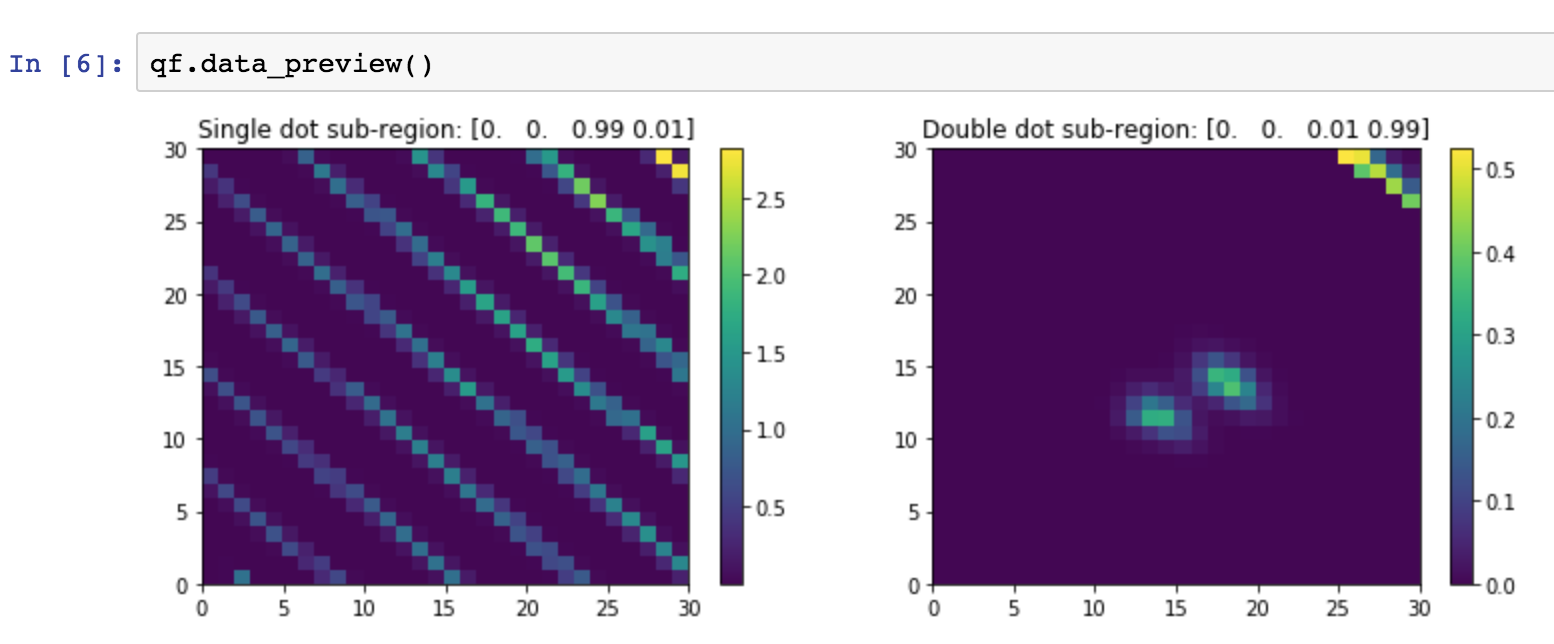}
\caption{{\bf A sample preview output.} 
A preview of a single dot (left) and of a double dot subregion (right) generated using a {\it QFlow lite} build-in function \texttt{qf.data\_preview()}. The labels printed above images indicate the actual state, as well as the fraction of each type of state within the given image in the format: [SC, QPC, SD, DD].
} 
\label{fig:ql-data-prev}
\end{figure}

After initializing the QFlow class that accompanies the {\it QFlow lite} suite (Step 1), the user can modify the size of the images that will be generated for training (Step 2). The total time to generate the training set from the 1001 stored primitives on a 2014 MacBook Pro is about 2.5 minutes (a default of 10 subregions per data file, $30\times30$ pixels each, and $10,010$ effective realizations). The size of the training images can be modified directly within the {\it QFlow lite} suite. The number of images generated from each device has to be changed within the class itself. Once the training set is established, the user can preview the data. A sample of a single and double dot, together with the label of each region, are generated. The label indicates the actual state, as well as the fraction of each type of state within the given image, and is in the format: [SC, QPC, SD, DD] (see Fig~\ref{fig:ql-data-prev} for a sample preview output).

Within Step 3: Definition of the learning network, the user can modify the configuration of the neural network and other parameters of the learning algorithm (directly within {\it QFlow lite}). Once the network architecture is set, the user can proceed to training (Step 4) and then to visualization (Step 5) of the accuracy of the learner. Steps 3 through 5 can be repeated until the desired accuracy is achieved. The final step (Step 6) allows for valuation of the trained network on experimental data.
 
\section*{Conclusion}
From hand-writing recognition of postal addresses~\cite{Denker89-ZIP} to defeating the champions of chess~\cite{IBM-chess} or the game of Go~\cite{Google-GO,Silver16-GO} to self-driving cars~\cite{self-drive} to Amazon and Netflix recommendations~\cite{amazon,netflix} to helping battle cancer~\cite{Montgomery16-AIFC,O'Hare17-AIBC}, machine learning algorithms are becoming an integral part of our daily lives. Over the past few years machine learning has also revolutionized how research and science are being done. `Adam'---an artificially-intelligent robot scientist---uses ML
algorithms to analyze parameters from multiple experimental replicates and helps identify genes responsible for catalyzing specific reactions in the metabolic pathways~\cite{Sparkes10-Adam}. Robot `Eve' uses machine learning to help with search for new drugs~\cite{Williams15-Eve}.

Bringing machine learning to automate discovery or other processes rather than using a `brute force' approach has the potential to substantially accelerate scientific progress across many disciplines. However, the success of ML algorithms depends on the quality of the training data. To enable research on quantum dot tuning automation, we developed a reliable dataset of simulated current and charge sensor response values versus gate voltage. Using this training set, we showed that the algorithm can identify the state of the quantum dots in both simulation and experiment, and then tune the device to a desired electronic state (see Ref.~\cite{Kalantre18-MTQ} for details about tuning and working with experimental data). This establishes a closed-loop system for tuning devices and eliminates the need for human intervention. To enable the usage and further study of ML for quantum dot devices, the training dataset is available via the Midas system at NIST and at data.gov~\cite{qf-data}.

The applications of quantum dots are plenty, from studying quantum phenomena observed in real systems~\cite{Ashoori96-EAA,Manouchehri08-QWD,Melnikov16-QWD} to micro-masers and light emitting diodes~\cite{Liu15-SDM} to realizing quantum bits~\cite{Loss98-QCD} to building quantum-dot-based single-photon sources~\cite{Unitt05-DPS,Muller18-LED}. Progress in quantum dots array sizes has been steady. Given that double and triple dots are being routinely used in experiments~\cite{vanDiepen18-ATC,Botzem2018,Baart2016} and moderate linear array sizes ($\sim$ 10 dots)~\cite{Zajac16-SGA}, as well as two-dimensional arrays~\cite{Mukhopadhyay18-2DD}, are on the horizon, an auto-tuning procedure for these devices is a significant step for employing them in both the laboratories and in applications.

While we focused on double dot devices, the procedure generalizes to higher dimensional systems and, hence, to tuning larger quantum dot arrays. We are currently working on a full version of the {\it QFlow} suite that will allow the user to choose the type of data to work with (i.e., the current or charge sensor measurements) as well as to automatically tune the device to a desired state. In the future, we intend to expand our approach to systems with a larger number of dots in both one as well as two dimensions. This should enable ML approaches to take over device control as experiments go from the routine few to the challenging many. The fruitful exchange between modern machine learning techniques and quantum technology thus presents an exciting and promising pathway for realizing scalable quantum dot devices for computing and other applications.

\section*{Supporting information}\label{sec:sup-inf}

\paragraph*{S1 File}\label{S1_file}
{\bf Data visualization code.} A Jupyter notebook-based code for converting the NumPy files to csv format and for previewing the simulated data.

\section*{Acknowledgments}
 The authors acknowledge the University of Maryland supercomputing resources (\url{http://hpcc.umd.edu}) made available for conducting the research reported in this paper. Any mention of commercial products is for information only; it does not imply recommendation or endorsement by NIST.

\nolinenumbers



\begin{thebibliography}{10}

\bibitem{Ladd10-QC} Ladd~TD, Jelezko~F, Laflamme~R, Nakamura~Y, Monroe~C, O’Brien~JL.
\newblock Quantum computers.
\newblock Nature 2010; 464: 45--53.

\bibitem{Nielsen11-QCI} Nielsen~MA, Chuang~IL.
\newblock Quantum Computation and Quantum Information. 10th ed.
\newblock New York, NY, USA: Cambridge University Press (2011).

\bibitem{Gambetta17-QSC} Gambetta~JM, Chow~JM, Steffen~M.
\newblock Building logical qubits in a superconducting quantum computing system.
\newblock npj Quantum Information 2017; 3: 2.

\bibitem{Kielpinski02-LIC} Kielpinski~D, Monroe~C, Wineland~DJ.
\newblock Architecture for a large-scale ion-trap quantum computer.
\newblock Nature 2002; 417: 709--711.

\bibitem{Lekitsch17-MTQ} Lekitsch~B, Weidt~S, Fowler~AG, M{\o}lmer~K, Devitt~SJ, Wunderlich~C, Hensinger~WK.
\newblock Blueprint for a microwave trapped ion quantum computer.
\newblock Sci Adv 2017; 3 (2): e1601540.

\bibitem{Mukhopadhyay18-2DD} Mukhopadhyay~U, Dehollain~JP, Reichl~C, Wegscheider~W, Vandersypen~LMK.
\newblock A 2$\times$2 quantum dot array with controllable inter-dot tunnel couplings.
\newblock Appl Phys Lett 2018; 112 (18): 183505.

\bibitem{Zajac16-SGA} Zajac~DM, Hazard~TM, Mi~X, Nielsen~E, Petta~JR.
\newblock Scalable Gate Architecture for a One-Dimensional Array of Semiconductor Spin Qubits.
\newblock Phys Rev Appl 2016; 6: 054013.

\bibitem{Li17-SQD} Li~R, Petit~L, Franke~DP, Dehollain~JP, Helsen~J, Steudtner~M, et~al.
\newblock A crossbar network for silicon quantum dot qubits.
\newblock Sci Adv 2018; 4 (7): eaar3960.

\bibitem{Karzig2017} Karzig~T, Knapp~C, Lutchyn~RM, Bonderson~P, Hastings~MB, Nayak~C, et~al.
\newblock Scalable designs for quasiparticle-poisoning-protected topological quantum computation with Majorana zero modes.
\newblock Phys Rev B 2017; 95 (23): 1--34.

\bibitem{Neill2017} Neill~C, Roushan~P, Kechedzhi~K, Boixo S, Isakov SV, Smelyanskiy V, et~al.
\newblock A blueprint for demonstrating quantum supremacy with superconducting qubits.
\newblock Science 2018; 360 (6385): 195--199.

\bibitem{Saffman2016} Saffman~M.
\newblock Quantum computing with atomic qubits and Rydberg interactions: Progress and challenges.
\newblock J Phys B: At Mol Opt Phys 2016; 49: 202001.

\bibitem{Sete2016} Sete~EA, Zeng~WJ, Rigetti~CT.
\newblock A functional architecture for scalable quantum computing.
\newblock In: 2016 IEEE International Conference on Rebooting Computing (ICRC); 2016 Oct 17-19; San Diego, CA. IEEE; 2016. pp. 1--6.

\bibitem{Blais2004} Blais~A, Huang~RS, Wallraff~A, Girvin~SM, Schoelkopf~RJ.
\newblock Cavity quantum electrodynamics for superconducting electrical circuits: An architecture for quantum computation.
\newblock Phys Rev A 2004; 69 (6): 1--14.

\bibitem{Brown2016} Brown~KR, Kim~J, Monroe~C.
\newblock Co-Designing a Scalable Quantum Computer with Trapped Atomic Ions.
\newblock npj Quantum Information 2016; 2: 16034.

\bibitem{Bernien17} Bernien~H, Schwartz~S, Keesling~A, Levine~H, Omran~A, Pichler~H, et~al.
\newblock Probing many-body dynamics on a 51-atom quantum simulator.
\newblock Nature 2017; 551: 579--584.

\bibitem{Hanson07-SFQ} Hanson~R, Kouwenhoven~LP, Petta~JR, Tarucha~S, Vandersypen~LMK.
\newblock Spins in few-electron quantum dots.
\newblock Rev Mod Phys 2007; 79: 1217--1265.

\bibitem{Baart2016} Baart~TA, Eendebak~PT, Reichl~C, Wegscheider~W, Vandersypen~LMK.
\newblock Computer-automated tuning of semiconductor double quantum dots into the single-electron regime.
\newblock Appl Phys Lett 2016; 108 (21): 1--9.

\bibitem{Botzem2018} Botzem~T, Shulman~MD, Foletti~S, Harvey~SP, Dial~OE, Bethke~P, et~al.
\newblock Tuning methods for semiconductor spin-qubits. Preprint.
\newblock Available from: arXiv:1801.03755v1 (2018).

\bibitem{vanDiepen18-ATC} van Diepen~CJ, Eendebak~PT, Buijtendorp~BT, Mukhopadhyay~U, Fujita~T, Reichl~C, et~al.
\newblock Automated tuning of inter-dot tunnel coupling in double quantum dots.
\newblock Appl Phys Lett 2018; 113: 033101.

\bibitem{Krizhevsky12-CCN} Krizhevsky~A, Sutskever~I, Hinton~GE.
\newblock ImageNet Classification with Deep Convolutional Neural Networks.
\newblock In: Pereira~F, Burges~CJC, Bottou~L, Weinberger~KQ, editors. Advances in neural information processing systems 25 (NIPS 2012); Stateline, NV. Red Hook, NY, USA: Curran Associates, Inc; 2012. pp. 1097--1105.

\bibitem{Wiel02-DQD} {van der Wiel}~WG, {De Franceschi}~S, Elzerman ~JM, Fujisawa~T, Tarucha~S, Kouwenhoven~LP.
\newblock Electron transport through double quantum dots.
\newblock Rev Mod Phys 2003; 75: 1--22.

\bibitem{Kalantre18-MTQ} Kalantre~SS, Zwolak~JP, Ragole~S, Wu~X, Zimmerman~NM, Stewart Jr.~MD, Taylor~JM.
\newblock Machine learning techniques for state recognition and auto-tuning in quantum dots. Preprint.
\newblock Available from: arXiv:1712.04914v2 (2017).

\bibitem{qf-data} {National Institute of Standards and Technology}. 
\newblock Quantum dot data for machine learning. 2018 [cited 2018 Apr 30]. 
\newblock Database: data.gov [Internet].
\newblock  Available from: \url{https://catalog.data.gov/dataset/quantum-dot-data-for-machine-learning}.

\bibitem{Griffiths99} Griffiths~DJ.
\newblock Introduction to electrodynamics. 3rd ed.
\newblock Upper Saddle River, NJ: Prentice-Hall, Inc (1999).

\bibitem{Lundqvist83} Lundqvist~S, March~NH, editors.
\newblock Theory of the Inhomogeneous Electron Gas. 1st ed.
\newblock New York, NY, USA: Springer US (1983).

\bibitem{Johnson05-PHD} Johnson~AC.
\newblock Charge Sensing and Spin Dynamics in GaAs Quantum Dots. Ph.D. Thesis.
\newblock Harvard University. 2005.
\newblock Available from: \url{https://qdev.nbi.ku.dk/student_theses/pdf_files/A_Johnson_thesis.pdf}.

\bibitem{scipy} Jones~E, Oliphant~TE, Peterson~P, et~al.
\newblock {SciPy}: Open source scientific tools for {Python}. 2001-- [cited 2018 Apr 30].
\newblock Available from: \url{http://www.scipy.org/}.

\bibitem{numpy} Oliphant~TE.
\newblock Guide to NumPy. 1st ed.
\newblock USA: Trelgol Publishing (2006).

\bibitem{numpy-performance} Sarkar~T. 
\newblock Why you should start using .npy file more often\dots 2018 Mar 20 [cited 2018 Jun 5].
\newblock In: Towards Data Science [Internet].
\newblock Available from: \url{https://towardsdatascience.com/why-you-should-start-using-npy-file-more-often-df2a13cc0161}.

\bibitem{Diederik14} Kingma~DP, Ba~J.
\newblock Adam: {A} Method for Stochastic Optimization. Preprint.
\newblock Available from: arXiv:14126980v9 (2014).

\bibitem{Chinchor92-F1} Chinchor~N.
\newblock MUC-4 Evaluation Metrics.
\newblock In: MUC4 '92: Proceedings of the 4th Conference on Message Understanding; 1992 Jun 16-18; McLean, VA. Stroudsburg, PA, USA: Association for Computational Linguistics; 1992. pp. 22--29.

\bibitem{jupyter} Kluyver~T, Ragan-Kelley~B, P{\'e}rez~F, Granger~B, Bussonnier~M, Frederic~J, et~al.
\newblock Jupyter Notebooks -- a publishing format for reproducible computational workflows.
\newblock In: Loizides F, Schmidt B, editors. Positioning and Power in Academic Publishing: Players, Agents and Agendas. IOS Press; 2016. pp. 87--90.

\bibitem{qf-lite} QFlow Team 
\newblock QFlow Lite. 2018 [cited 2018 Apr 30].
\newblock Database: GitHub [Internet].
\newblock Available from: \url{https://github.com/jpzwolak/QFlow-lite}.

\bibitem{Denker89-ZIP} Denker~JS, Gardner~WR, Graf~HP, Henderson~D, Howard~RE, Hubbard~W, et~al.
\newblock Neural network recognizer for hand-written zip code digits.
\newblock In: Touretzky DS, editor. Advances in neural information processing systems 1 (NIPS 1988); Denver, CO. San Francisco, CA, USA: Morgan Kaufmann Publishers Inc; 1989. pp. 323--331.

\bibitem{IBM-chess}
\newblock Deep Blue. 2011 Sep 13 [cited 2018 May 31].
\newblock In: IBM’s 100 Icons of Progress [Internet].
\newblock Available from: \url{http://www-03.ibm.com/ibm/history/ibm100/us/en/icons/deepblue/}.

\bibitem{Google-GO} Gibney~E.
\newblock Google {AI} algorithm masters ancient game of Go.
\newblock Nature 2016; 529: 445--446.

\bibitem{Silver16-GO} Silver~D, Huang~A, Maddison~CJ, Guez~A, Sifre~L, {van den Driessche}~G, et~al.
\newblock Mastering the game of Go with deep neural networks and tree search.
\newblock Nature 2016; 529: 484--489.

\bibitem{self-drive} Jochem~T, Pomerleau~D, Kumar~B, Armstrong~J.
\newblock PANS: a portable navigation platform.
\newblock In: Proceedings of the Intelligent Vehicles '95 Symposium; 1995 Sep 25-26; Detroit, MI. IEEE; 1995. pp. 107--112.

\bibitem{amazon} Levy~S.
\newblock Inside Amazon's Artificial Intelligence Flywheel. 2018 Jan 2 [cited 2018 Jun 06].
\newblock Wired [Internet].
\newblock Available from: \url{https://www.wired.com/story/amazon-artificial-intelligence-flywheel/}.

\bibitem{netflix} Ekanadham~C.
\newblock Using Machine Learning to Improve Streaming Quality at Netflix. 2018 Mar 22 [cited 2018 Jun 06].
\newblock In: Netflix Technology Blog [Internet].
\newblock Available from: \url{https://medium.com/netflix-techblog/using-machine-learning-to-improve-streaming-quality-at-netflix-9651263ef09f}.

\bibitem{Montgomery16-AIFC} Montgomery~M.
\newblock In Cancer Fight, Artificial Intelligence Is A Smart Move For Everyone. 2016 Dec 22 [cited 2018 May 31].
\newblock In: Forbes [Internet].
\newblock Available from: \url{https://www.forbes.com/sites/mikemontgomery/2016/12/22/in-cancer-fight-artificial-intelligence-is-a-smart-move-for-everyone/#33c4487e4064}.

\bibitem{O'Hare17-AIBC} O'Hare~R.
\newblock Research collaboration aims to improve breast cancer diagnosis using AI. 2017 Nov 24 [cited 2018 May 31].
\newblock In: Imperial College London [Internet].
\newblock Available from: \url{https://www.imperial.ac.uk/news/183293/research-collaboration-aims-improve-breast-cancer/}.

\bibitem{Sparkes10-Adam} Sparkes~A, Aubrey~W, Byrne~E, Clare~A, Khan~MN, Liakata~M, et~al.
\newblock Towards Robot Scientists for autonomous scientific discovery.
\newblock Autom Exp 2010; 2: 1.

\bibitem{Williams15-Eve} Williams~K, Bilsland~E, Sparkes~A, Aubrey~W, Young~M, Soldatova~LN, et~al.
\newblock Cheaper faster drug development validated by the repositioning of drugs against neglected tropical diseases.
\newblock J R Soc Interface 2015; 12: 20141289.

\bibitem{Ashoori96-EAA} Ashoori~RC.
\newblock Electrons in artificial atoms.
\newblock Nature 1996; 379: 413--419.

\bibitem{Manouchehri08-QWD} Manouchehri~K, Wang~JB.
\newblock Quantum walks in an array of quantum dots.
\newblock J Phys A: Math Theor 2008; 41: 065304.

\bibitem{Melnikov16-QWD} Melnikov~AA, Fedichkin~LE.
\newblock Quantum walks of interacting fermions on a cycle graph.
\newblock Sci Rep 2016; 6: 34226.

\bibitem{Liu15-SDM} Liu~YY, Stehlik~J, Eichler~C, Gullans~MJ, Taylor~JM, Petta~JR.
\newblock Semiconductor double quantum dot micromaser.
\newblock Science 2015; 347 (6219): 285--287.

\bibitem{Loss98-QCD} Loss~D, DiVincenzo~DP.
\newblock Quantum computation with quantum dots.
\newblock Phys Rev A 1998; 57 (1): 120.

\bibitem{Unitt05-DPS} Unitt~DC, Bennett~AJ, Atkinson~P, Cooper~K, See~P, et~al.
\newblock Quantum dots as single-photon sources for quantum information processing.
\newblock J. Opt. B: Quantum Semiclass. Opt. 2005; 7: S129.

\bibitem{Muller18-LED} M\"uller~T, Skiba-Szymanska~J, Krysa~AB, Huwer~J, Felle~M, et~al.
\newblock A quantum light-emitting diode for the standard telecom window around 1,550~nm.
\newblock Nat Commun 2018; 9: 862.

\end{thebibliography}
\end{document}